\documentclass[english,a4paper,twocolumn,amsmath,amssymb]{revtex4-1}
\usepackage{amsmath}
\usepackage{amssymb}
\usepackage{graphicx,color,soul}
\usepackage[utf8]{inputenc}

\def\mydoubleq#1{``#1''}
\def\mysingleq#1{`#1'}

\begin{document}
	
\title{Relaxation in a Phase-separating Two-dimensional Active Matter System with Alignment Interaction}
	
\author{Saikat Chakraborty$^{1,2}$ and Subir K. Das$^{1,3,}$} 
\email[]{das@jncasr.ac.in}
\affiliation{$^1$Theoretical Sciences Unit, Jawaharlal Nehru Centre for Advanced Scientific Research, Jakkur P.O., Bangalore 560064, India}
\affiliation{$^2$Institut f\"{u}r Physik, Johannes-Gutenberg Universit\"{a}t Mainz, Staudinger Weg 7-9, 55128 Mainz, Germany}
\affiliation{$^3$School of Advanced Materials, Jawaharlal Nehru Centre for Advanced Scientific Research, Jakkur P.O., Bangalore 560064, India}

\begin{abstract}
{Via computer simulations we study kinetics of pattern formation in a two-dimensional active matter system. Self-propulsion in our model is incorporated via the Vicsek-like activity, i.e., particles have the tendency of aligning their velocities with the average directions of motion of their neighbors. In addition to this dynamic or active interaction, there exists passive inter-particle interaction in the model for which we have chosen the standard Lennard-Jones form. Following quenches of homogeneous configurations to a point deep inside the region of coexistence between high and low density phases, as the systems exhibit formation and  evolution of particle-rich clusters, we investigate properties related to the morphology, growth and aging. A focus of our study is on the understanding of the effects of structure on growth and aging. To quantify the latter we use the two-time order-parameter autocorrelation function. This correlation, as well as the growth, is observed to follow power-law time dependence, qualitatively similar to the scaling behavior reported for passive systems. The values of the exponents have been estimated and discussed by comparing with the previously obtained numbers for other dimensions as well as with the new results for the passive limit of the considered model. We have also presented results on the effects of temperature on the activity mediated phase separation.}
\end{abstract}

\maketitle

\section{Introduction}
Active matter systems are composed of self-propelled particles and are inherently non-equilibrium in nature~\cite{vicsek1995, toner1995, greg2004, vicsek2012, marchetti2013, elgeti2015, cates2015, chate2020}. These systems are of significant recent interest~\cite{baurle2018,partridge2019,mahault2019,agudo2019,peruani2019,foffano2019,maitra2020,reigh2020,wagner2019,toneian2019}, particularly in connection with the clustering phenomena that are commonly observed in a variety of systems containing living entities. Examples of such systems are aplenty, bacterial colony~\cite{berg2005}, school of fish~\cite{lopez2012} and flock of birds~\cite{cavagna2014} being some of the popular ones. 

In one of the physically relevant models, constructed to understand such clustering, referred to as the Vicsek model~\cite{vicsek1995}, one considers alignment of velocities of particles along the average directions of motion in the neighborhoods. There, of course, exists non-alignment interactions in many real active matter systems~\cite{howse2007, palacci2013, tenhagen2011, redner2013, basu2018, tailleur2008}. Nevertheless, the Vicsek model remains~\cite{das2014, trefz2016, trefz2017, liebchen2017} a  paradigm for studying phase transitions~\cite{fisher1967} in this domain, with strong relevance to clustering in a plethora of open biological systems~\cite{ballerini2008,nagy2010,klumpp2019}. 

In the statistical mechanics community there has been significant drive in terms of developing  methodologies and constructing concepts with the purpose of understanding these inherently driven systems. Large fraction of these works have connection with phase transitions. A sub-domain in the area of phase transitions is concerned about the universality classes associated with the scaling properties of various dynamic quantities~\cite{hohenberg1977, bray1995, dfisher1988, yeung1996, mcleish2003, das2006,roy2011,midya2017jcp}. In the context of clustering or coarsening dynamics, the important aspects are nonequlibrium morphology, growth and aging~\cite{bray1995,dfisher1988,yeung1996,mcleish2003,midya2017,shimizu2015,majumder2010,testard2014,midya2015,roy2019}. In passive systems related scaling properties and corresponding universality classes are dependent upon~\cite{bray1995} symmetry and conservation of order-parameter, space dimension ($d$), transport mechanism, etc. 

In spite of certain similarities between clustering in active and passive matter systems, above  properties in the former, particularly the aging phenomena, has only been occasionally  investigated~\cite{janssen2017,das2017,chen2012,sokolov2012}. In this article we undertake the task of studying aging in a Vicsek-like model in $d=2$, in conjunction with morphology and growth. By comparing these results with the existing literature, our intention is to obtain information on the influence of activity on coarsening phenomena and obtain better understanding on the connection among the above mentioned properties.  

Often aging is studied via the two-time order-parameter autocorrelation function~\cite{dfisher1988,roy2019}
\begin{equation}\label{eq:auto}
C_{\textrm{ag}}(t,t_{w})=\langle{\psi(\vec{r},t) \psi(\vec{r},t_{w})}\rangle -\langle{\psi(\vec{r},t)}\rangle \langle{\psi(\vec{r},t_{w})}\rangle,
\end{equation}
where $t$ and $t_w$ ($\le t$) are observation and waiting times, respectively. The latter time is also referred to as the age of a growing system. In Eq.~\eqref{eq:auto}, $\psi(\vec{r},t)$ is an appropriate space ($\vec{r}$) and time dependent order parameter, which is a scalar quantity for the present problem. In a wide range of passive systems, $C_{\textrm{ag}}(t,t_w)$ exhibits the scaling behavior~\cite{dfisher1988}
\begin{eqnarray}\label{eq:cagscl}
C_{\textrm{ag}}(t,t_{w}) \equiv \tilde{C}_{\textrm{ag}} (\ell/\ell_w).
\end{eqnarray}
In Eq.~\eqref{eq:cagscl}, $\ell$ and $\ell_w$ are the characteristic lengths of the system at times $t$ and $t_w$, respectively. 

The characteristic length, i.e., the average size of domains, rich and poor in particles of a specific type, in a growing system often shows power-law scaling~\cite{bray1995}, i.e.,
\begin{equation}\label{eq:growth}
\ell \sim t^{\alpha}, 
\end{equation}
where $\alpha$ is the growth exponent. In many situations, $C_{\textrm{ag}}$ also shows power-law decay of the form~\cite{dfisher1988}
\begin{equation}\label{eq:lambda}
C_{\textrm{ag}} \sim (\ell/\ell_w)^{-\lambda},
\end{equation}
$\lambda$ being referred to as the aging exponent. The values of $\alpha$ and $\lambda$ are expected to depend crucially on the transport mechanism and morphology~\cite{siggia1979, binder1976,roy2019, roy2013}. It has been argued, if in the limit of small wave number ($k$) the structure factor has the power-law behavior~\cite{yeung1988} $S(k,t) \sim k^\beta$, $\lambda$ should obey the bound~\cite{yeung1996} 
\begin{equation}\label{eq:yrd}
\lambda \geq \frac{d+\beta}{2}.
\end{equation}

Our objective in this work is to investigate the influence of alignment activity on the values of $\alpha$ and $\lambda$, as well as on morphology, in a phase-separating system where one of the phases has solid-like arrangement of particles. For this purpose we perform (hybrid) molecular dynamics (MD) simulations~\cite{allen1987,frenkel1996} of a two-dimensional system of active particles having an  inter-particle passive interaction and Vicsek-like~\cite{vicsek1995} self-propulsion. The value of the aging exponent $\lambda$, as well as the growth exponent $\alpha$,  are estimated accurately for the considered model. We examine whether the obtained value of $\lambda$ is consistent with the bound in Eq.~\eqref{eq:yrd}. Discussions by comparing our active matter results with the newly obtained ones for the passive limit of the model, as well as with those, obtained previously~\cite{das2017}, for active matter systems in other dimensions, wherever appropriate, are provided. We have also demonstrated the effects of temperature on activity mediated phase separation.

In passive systems it has been observed that the exponents associated with coarsening phenomena, like in equilibrium critical phenomena, are often decided by the space dimension. E.g., the aging exponent for Ising model depends strongly on dimensions. In Ref.~\cite{das2017} we had undertaken study of phase separation in $d=3$ for the model considered here. Keeping the picture of passive systems in mind, our study in $d=2$ has an objective of understanding dimensionality dependence of coarsening in active matter systems.

The rest of the paper is organized as follows. In section \ref{sec:method} we describe the model and methods. Results are presented in section \ref{sec:result}. Finally, we conclude the paper in section \ref{sec:conclusion} by presenting a summary.
\section{Model and Methods}\label{sec:method}
In our system the constituent particles interact with each other via the pair-potential (for $r<r_c$)~\cite{allen1987}:
\begin{equation}\label{eq:pairpot}
u(r)=U(r)-U(r_{c})-(r-r_{c})\bigg(\frac{dU}{dr}\bigg)_{r=r_{c}},
\end{equation}
where $r$ is the scalar distance between two particles and $U(r)$ is the standard passive interaction of Lennard-Jones (LJ) form~\cite{allen1987,frenkel1996}, i.e., 
\begin{equation}\label{eq:lj}
U(r)=4\epsilon \bigg[\bigg(\frac{\sigma}{r}\bigg)^{12}-\bigg(\frac{\sigma}{r}\bigg)^{6}\bigg].
\end{equation}
In Eq.~\eqref{eq:lj} $\epsilon$ and $\sigma$ are interaction strength and particle diameter, respectively, while $r_c$ ($=2.5\sigma$), in Eq.~\eqref{eq:pairpot}, is a cut-off distance that is introduced for faster computation. Given that the LJ potential is a short-range one, introduction of $r_c$ does not alter the critical universality class~\cite{fisher1967}. After the abrupt cut and shift [see the second term on the right hand side], the introduction of the third term in Eq.~\eqref{eq:pairpot} becomes necessary to make the force continuous at $r=r_c$. The form of the force correction term is, however, not unique~\cite{allen1987,voigtmann2009}. 

For this model, the critical  values of temperature and number density, for a vapor-liquid transition~\cite{midya2017jcp2}, are $T_c \simeq 0.41 \,{\epsilon}/{k_B}$ and $\rho_c \simeq 0.37$, in $d=2$, where $k_B$ is the Boltzmann constant. For a simulation box of area $A$, containing $N$ particles, the particle density is defined as $\rho=N/A$. We have considered square boxes of linear dimension $L \sigma$. 

The self-propulsion in the system is invoked by following the well-known Vicsek activity~\cite{vicsek1995}, as already  mentioned. Here, the direction of a particle's motion is influenced by that of its neighbors. This is implemented in the following manner~\cite{das2017}. At each time step, an external force, strength of which is given by a parameter $f_A$, taken to be unity throughout for the active case, is imparted on each particle. The direction of the force is along the average velocity of all the particles contained within the radius of neighborhood $R_n$ of the concerned particle, the value of which, in this work, we set to $r_c$. Following this exercise the magnitudes of the velocities of the particles were restored to the original values. Thus, like in the original Vicsek model~\cite{vicsek1995}, this whole procedure changes only the directions of motion of the particles.   

We performed MD simulations~\cite{allen1987,frenkel1996} for the passive interactions. The temperature ($T$) during the simulations was controlled via the Langevin thermostat~\cite{das2017, allen1987, frenkel1996}. This implies, for particle $i$ we have solved the equation
\begin{equation}\label{eq:lngvn}
m\ddot{\vec{r}}_{i}=-\vec{\nabla}{u_{i}}-m\gamma \dot{\vec{r}}_{i}+\sqrt{2m\gamma k_{B}T} \vec{R}(t).
\end{equation}
Here, $m$ is the mass, same for all particles, $u_i$ is the energy of the $i^{\text{th}}$ particle originating from the inter-particle potential, $\gamma$ is a damping constant and $\vec{R}(t)$ is a noise having delta correlation in space and time. We have used the Verlet velocity algorithm to solve Eq. \eqref{eq:lngvn}, with time step size $\Delta t=0.005$, in units of $\sqrt{{m{\sigma^2}}/{\epsilon}}$. For the rest of the paper $m$, $\epsilon$, $\sigma$, $\gamma$ and $k_B$ have been set to unity.

For the purpose of analysis, the original configurations are mapped onto those on a square lattice~\cite{majumder2011,roy2013} with grids of linear dimension $\sigma$. To a lattice site we have assigned an order-parameter value $\psi=-1$, if it is empty, otherwise we have put $\psi=+1$. For quantifying the decay of $C_{\textrm{ag}}(t,t_w)$, as well as for independent understanding, the characteristic length was calculated as the distance at which the two-point equal time correlation function~\cite{bray1995},
\begin{equation}\label{eq:cor}
C(r,t) = \langle{\psi(\vec{r},t) \psi(\vec{0},t)}\rangle -\langle{\psi(\vec{r},t)}\rangle \langle{\psi(\vec{0},t)}\rangle,
\end{equation}
with $r$ ($=|\vec{r}|$) being the scalar separation between two lattice sites, decays to $0.2$ times its maximum value. Note that in order to obtain a meaningful estimation of growth and aging dynamics, it is necessary that $C(r,t)$ satisfies the scaling property~\cite{bray1995}
\begin{equation}\label{eq:crscl}
C(r,t) = \tilde{C}\bigg(\frac{r}{\ell(t)}\bigg).
\end{equation}

 From random initial configurations, in position as well as in velocity, the systems were primarily quenched to $T=0.25$, with $\rho=0.37$. As mentioned above, we have presented results from quenches to other values of $T$ as well. We have applied periodic boundary conditions in both the directions. Quantitative results are presented after averaging over about $60$ independent initial realizations, unless otherwise mentioned. At the above mentioned temperature, in the passive limit of the model we have observed solid-like  order~\cite{midya2017}. It remains to be seen if self-propulsion can alter the regular structure. If the basic structure remains the same, our objective, i.e., of studying the influence of activity on phase separation in systems with one of the phases being solid-like, will be met. Unless otherwise mentioned all our results are from $T=0.25$.
\begin{figure}[h!]
	\centering
	\includegraphics*{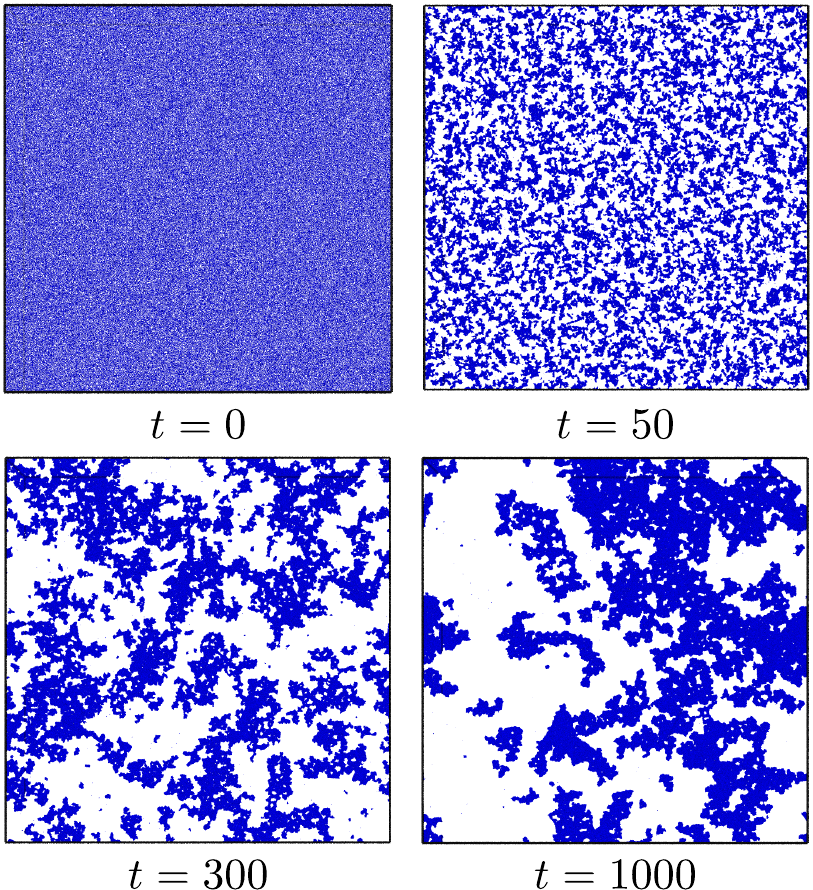}
	\caption{Snapshots, taken during the evolution of the considered active matter model at $T=0.25$, are shown from four different times. Particle coordinates are marked. These and subsequent results for nonzero activity, if not mentioned otherwise, are obtained from simulations with $L=1024$.
	}
	\label{fig:snap}
\end{figure}
\begin{figure}[h!]
	\centering
	\includegraphics*{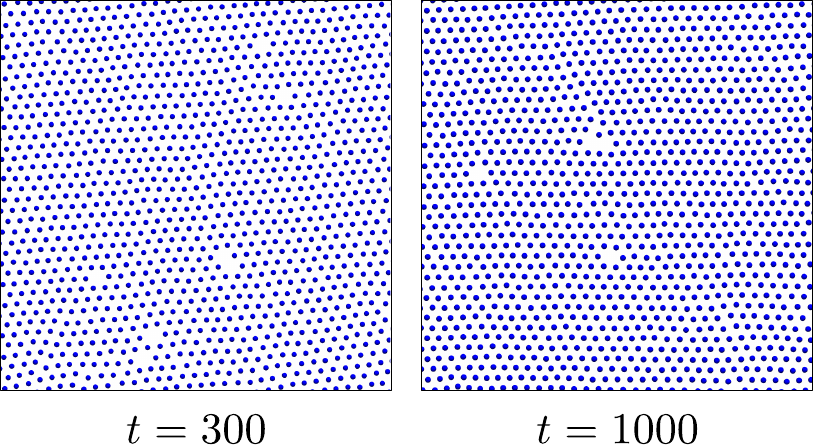}
	\caption{Enlarged views of parts of clusters from the snapshots at $t=300$ and $1000$ in Fig.~\ref{fig:snap}.	
	}
	\label{fig:cryst}
\end{figure}
\section{Results}\label{sec:result}
Time evolution of the model active matter system, starting from a homogeneous configuration, has been demonstrated in Fig.~\ref{fig:snap}. Following the quench to the final temperature, i.e., $T=0.25$, the system falls unstable to fluctuations. This leads to interesting pattern formation where particle-poor and particle-rich domains coexist with each other. With passing time this pattern grows. It is noticeable from the snapshots in Fig.~\ref{fig:snap} that there exists a fair degree of departure of the pattern from the standard bicontinuous morphology that is observed during phase separation in the passive systems, say in the Ising lattice-gas model~\cite{landau2005}, or for $f_A=0$ in the present model~\cite{midya2020}, with such high particle density.

In the passive limit of the model, as already mentioned, at such low temperature fairly regular arrangement of the particles is observed in the cluster regions~\cite{midya2017}. Such short-range solid-like order is being observed even when the activity is turned on. This can be appreciated from  Fig.~\ref{fig:cryst} where we have shown small parts of the late time snapshots that are presented in Fig.~\ref{fig:snap}. This implies that our study is related to phase separation in active matter with the high density phase being \mydoubleq{solid}. We will provide more quantitative information on this aspect soon. This and nearly connected structure of domains imply that hydrodynamic mechanism~\cite{hohenberg1977,midya2017,shimizu2015,siggia1979,binder1976,roy2013} is less important and thus, the adopted simulation method is physical. The solid-like order, however, like in many real (passive) systems, keeps changing from region to region. In $d=2$, of course, we do not expect very long range order. In our understanding there exist studies on coarsening in active fluids. Here the key question we ask: How does the growth dynamics get changed due to introduction of activity when one of the phases is solid?

\begin{figure}[h!]
	\centering
	\includegraphics*{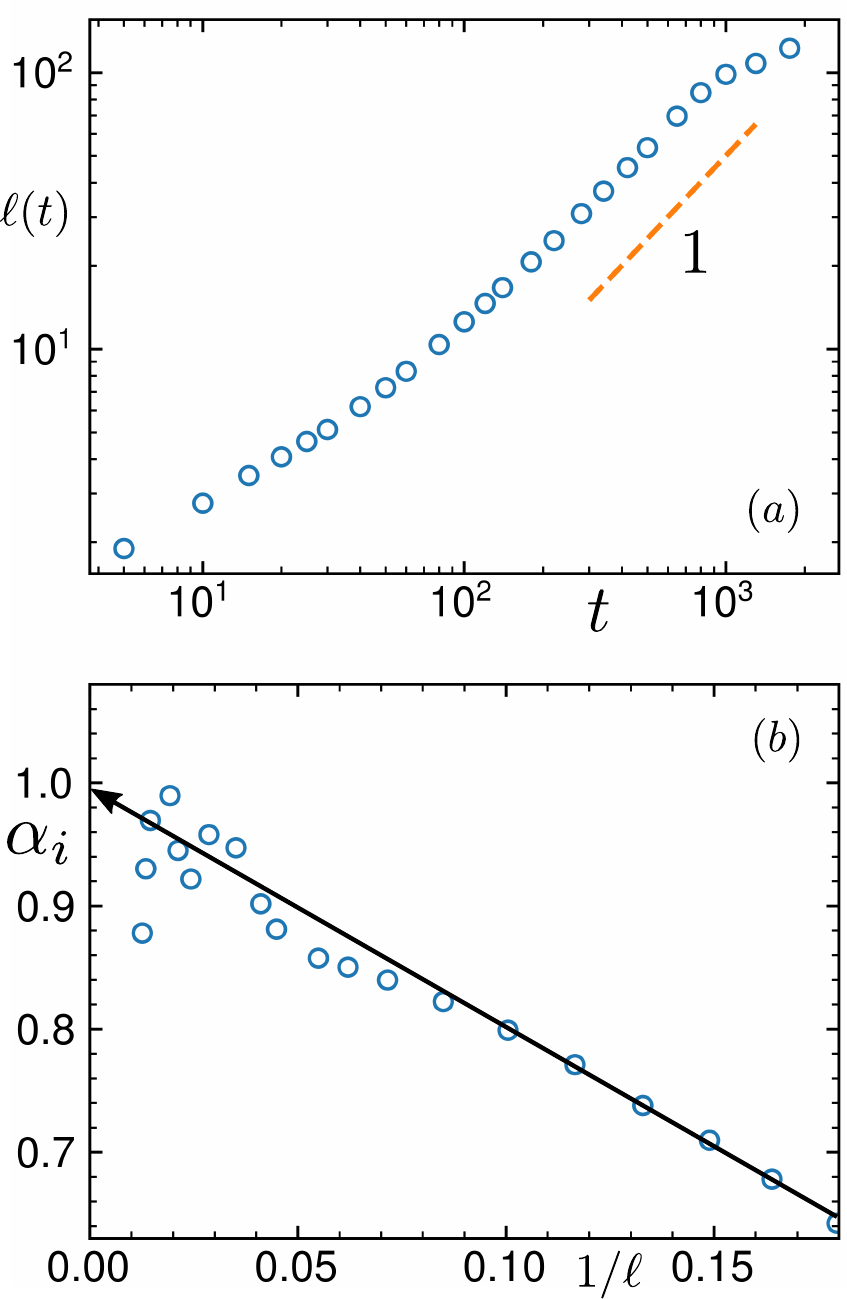}
	\caption{(a)~Average cluster length, $\ell(t)$, is plotted as a function of time, on a double-log scale. The dashed line represents a power law with exponent $1$. 
		(b)~Instantaneous exponent (running averaged), $\alpha_i$, for the growth data in (a), is plotted as a function of $1/\ell$. The solid line, with arrow-head, is a linear guide to the limit $\ell = \infty$.
	}
	\label{fig:growth}
\end{figure}
As mentioned above, for the purpose of analyses we have mapped the original configurations onto a square lattice. From Fig.~\ref{fig:cryst}, however, it is clear that the original configurations contain other regular structure(s). Nevertheless, this keeps the objective valid.

For quantitative understanding of growth, in Fig.~\ref{fig:growth}(a) we show the plot of $\ell(t)$ as a function of $t$, on a log-log scale, over a time range twice as large as three decades. At late time the data appear to be consistent with a power-law behavior with exponent $1$. Nevertheless, given that there is continuous curvature in the plot, further analysis is essential to arrive at a correct conclusion on the value of the exponent. In such a situation, in the literature of kinetics of phase transitions, there exists the tradition of calculating the time-dependent or instantaneous exponent~\cite{huse1986,majumder2010} 
\begin{equation}
\alpha_i = \frac{d \ln \ell}{d \ln t}.
\end{equation}
In Fig.~\ref{fig:growth}(b) we have presented a plot of $\alpha_i$ versus $1/\ell$. It is clearly seen that in the limit $\ell = \infty$, the convergence is towards $\alpha=1$, to a good degree of accuracy. 
Such instantaneous exponents are typically very noisy. However, here the results are reasonably smooth. This is due to good statistics.

The inconsistency of the early time data, in the direct log-log plot, with this value can be due to the presence of significant off-set (non-zero initial value of $\ell$)~\cite{majumder2010,amar1988} in the data set, as well as due to early time corrections or crossover. Note that the average cluster size even for a random spatial configuration is nonzero. This fact may lead to incorrect conclusion, if drawn from the appearance in a log-log plot, when the ranges of time and length scales are not large enough. Presence of normal curvature in domain boundaries, as well as that due to interface roughening~\cite{corberi2008}, can also lead to bending in the plot at early time, implying finite-time corrections to the asymptotic behavior~\cite{huse1986}. Bending can also be~\cite{bray1995} there due to different leading mechanisms of growth in different time regimes giving rise to crossovers in the values of $\alpha$. 

When the bending is entirely due to the off-set, one expects~\cite{amar1988,majumder2010}
\begin{equation}
\alpha_i=\alpha\bigg[1-\frac{\ell_0}{\ell}\bigg],
\end{equation}
where $\ell_0$ is the average cluster size at $t=0$. Given that we have obtained $\alpha=1$, a linear behavior with a slope \mysingleq{$-\ell_0$} in Fig.~\ref{fig:growth}(b) may suggest the absence of corrections or different growth regimes with crossovers. However, for the adopted method for estimating length, $\ell_0$ appears to be less than unity and this does not comply with the observed slope, implying the possibility of either corrections or crossovers. The source of any of these can be progressive solid-like arrangement of particles within the clusters, alongside other reasons that we are currently investigating. 

\begin{figure}[h!]
	\centering
	\includegraphics*{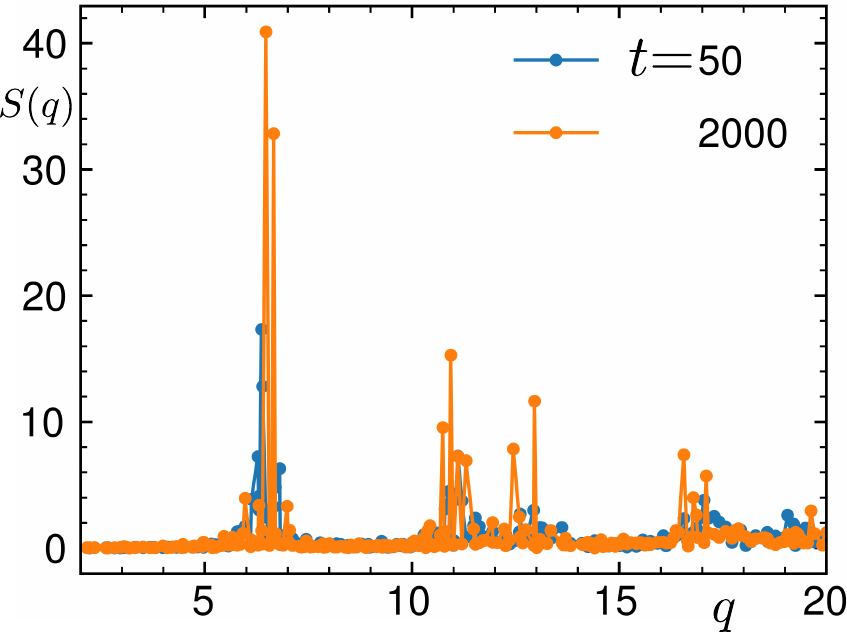}
	\caption{Structure functions $S(q)$, defined as $S(q) = \frac{1}{N} \langle \sum_{i=1,j=1}^{N} {\exp(i \vec{q} \cdot \vec{r})} \rangle$; $\vec{r}=\vec{r}_i-\vec{r}_j$, $N$ being the number of particles in the chosen part of the snapshot, from the continuum configurations, are plotted versus $q$. Results from two different times are shown. Only clustered regions were used for this calculation. Here we have used $q$, instead of $k$, for the wave number, in order to distinguish the objective from that of phase separation, the results for the latter being presented later.
	}
	\label{fig:hex}
\end{figure}

In Fig.~\ref{fig:hex} we have shown structure function~\cite{hansen2008,digregorio2018} $S(q)$ versus $q$ plots for the continuum configurations, calculated by using only clustered regions, like in Ref.~\cite{digregorio2018}. Results are shown from two different times. Regular arrangement of particles and enhancement in the order with the increase of time can be appreciated.  See captions for the definition of $S(q)$.

This growth of domains, like in the systems with hydrodynamics~\cite{siggia1979,binder1976,roy2013},  is very rapid. This is perhaps because of the fact that the cooperative motion, that the Vicsek alignment interaction leads to, has similarity with the advective flow in hydrodynamic environment. In fact, in $d=3$ the exponent~\cite{das2017} is the same as that for the viscous hydrodynamic growth in fluids~\cite{siggia1979,binder1976,roy2013}. In the present dimension, however, the growth is even stronger. Note that in $d=2$, the late time exponent~\cite{midya2017jcp2}, due to various hydrodynamic mechanisms, are expected to have the values $1/2$ and $2/3$.

\begin{figure}[h!]
	\centering
	\includegraphics*{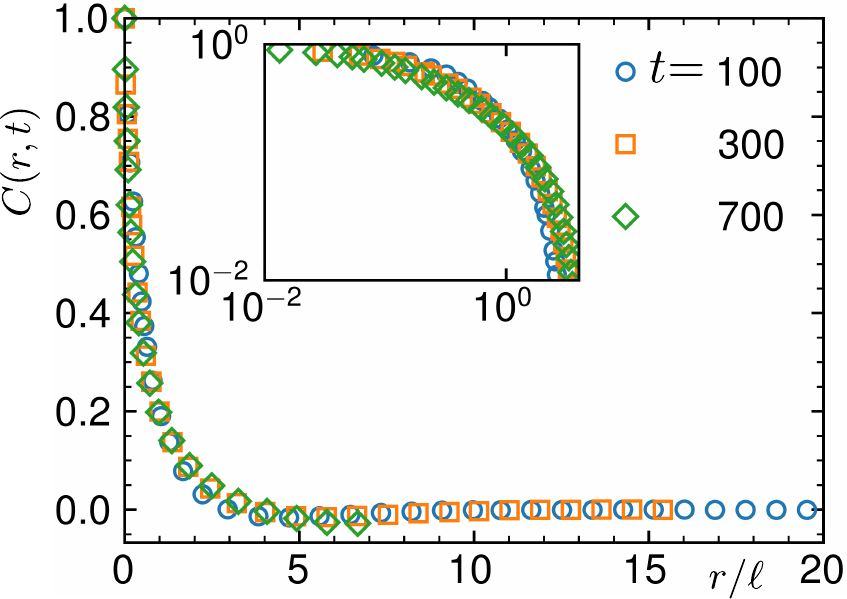}
	\caption{Plots of two-point equal time correlation function, $C(r,t)$, versus $r/\ell$. Data from three different times have been shown. The data sets are normalized in such a way that $C(0,t)=1$.
	Inset shows the same data sets in double-log scale.
	}
	\label{fig:sclcr}
\end{figure}
Above conclusion on a power-law growth will be meaningful if the structural self-similarity is present~\cite{bray1995} [cf. Eq.~\eqref{eq:crscl}]. To demonstrate that, in Fig.~\ref{fig:sclcr} we show plots of $C(r,t)$ from a few different times. In this exercise we have scaled the distance axis by $\ell$ and chosen the times from a range over which the log-log plot of $\ell$ versus $t$ data in Fig.~\ref{fig:growth}(a) appears reasonably consistent with the power-law exponent $\alpha=1$, thereby discarding the transient period as best as possible. 
\begin{figure}[!]
	\centering
	\includegraphics*{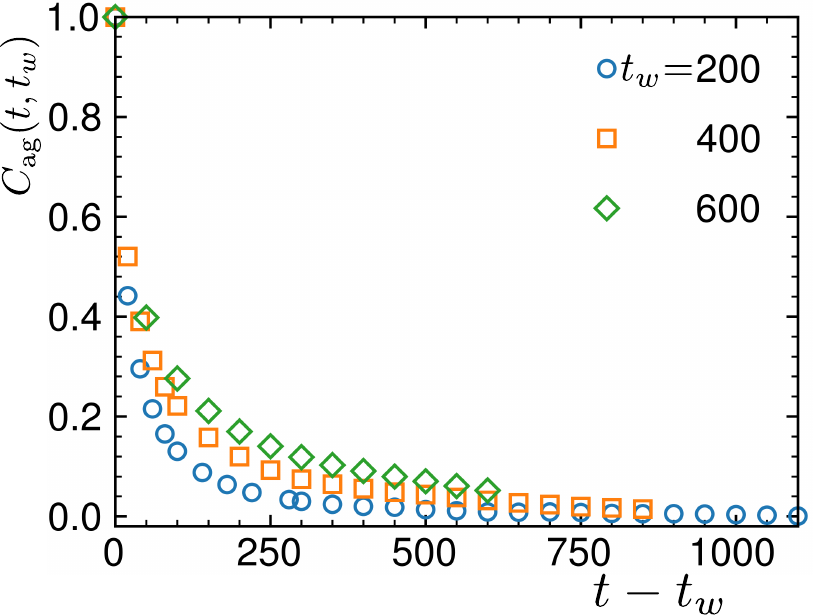}
	\caption{~Plots of the order-parameter autocorrelation function, $C_{\textrm{ag}}(t,t_w)$, versus $(t-t_w)$. Data for three waiting times are included. Like in Fig.~\ref{fig:sclcr}, here we have set $C_{\textrm{ag}}(t_w,t_w)=1$.
	}
	\label{fig:aging}
\end{figure}

The values of $t$ for the scaling plots of $C(r,t)$ cover a range reasonably close to a decade. For growth as fast as in the present case, it is difficult to illustrate the scaling over a very wide range of time, due to early emergence of the finite-size effects [see Fig.~\ref{fig:growth}(a)]. Consideration of $L$ twice as large, i.e., increase in the number of particles by a factor of four, will raise the time window by only a factor of two~\cite{das2012}, before the finite-size effects appear, while making the simulations extremely memory intensive. For slower dynamics, however, e.g., in Ising model with Kawasaki exchange kinetics~\cite{midya2015}, one includes data from a rather wide range of time. Important point to note here is that in the latter example, the range of length is relatively small~\cite{midya2015}. This will be demonstrated later when we present results for the passive limit of the present model. The dynamics in the passive limit is somewhat similar to the Kawasaki Ising model (KIM).

Nice collapse of data, despite some degree of fractality in the morphology,  except at very small $r/\ell$ regime, from all the presented times demonstrates self-similar growth. To emphasize on the quality of collapse we have presented the data on a log-log scale as well (see the inset of the Fig.~\ref{fig:sclcr}). It is worth mentioning here that a nonoverlapping outcome will be indicative of continuously changing character of morphology. In that case values of $\ell$ at different times, obtained from the method described in the previous section, will not provide information about the true growth in the system. Minor departure from collapse in small $r/\ell$ regime, we believe, is due to structural rearrangement at small length scale owing to the progress to solid-like order that is shown in Fig.~\ref{fig:hex}, which in turn can cause changes in interface roughening. We also believe that the solid-like order is responsible for the observed fractality in structure. If the corresponding dimension ($d_f$) is very low, it is instructive to introduce~\cite{vicsek1994} a factor $r^{d-d_f}$ in the scaling relation of Eq.~\eqref{eq:crscl}. Overall good collapse of the data sets indicate that $d_f$ is quite close to $d$. We will provide further discussion on this later.
 
The shape of the correlation function as well as the rate of growth differ from analogous situation in standard passive systems.  This is consistent with our discussion in the context of Fig.~\ref{fig:snap}. In the analogous passive situation $C(r,t)$ exhibits strong oscillations around the value \mydoubleq{$0$} for the presented range of the abscissa variable.
Given that we did not introduce hydrodynamics, in the passive limit it is expected that $\alpha$ should be close to $1/3$, the latter number being due to the Lifshitz-Slyozov mechanism~\cite{majumder2010,landau2005,huse1986,lifshitz1961}. This is valid for diffusive transport of material, as occurs in multi-component solid mixtures. Presence of activity, thus, in addition to modifying the morphology, significantly enhances the growth. Next, we move to quantify the aging dynamics using $C_{\textrm{ag}}(t,t_w)$.
 
In Fig.~\ref{fig:aging} we have plotted $C_{\textrm{ag}}(t,t_w)$ as a function of the translated time $(t-t_w)$. Data for different values of $t_w$ do not collapse on top of each other. Such deviation from time translation invariance is a characteristic feature of evolving, out-of-equilibrium systems. The shift in decay is monotonic, i.e., older the system (larger $t_w$) gets, slower the relaxation becomes. Like in the case of $C(r,t)$, here also, for the same reason, the range of $t_w$ is not very large. However, we emphasize again, the range of $\ell_w$ is comparable with or even larger than most of the studies with KIM~\cite{midya2015}. 
\begin{figure}[!]
	\centering
	\includegraphics*{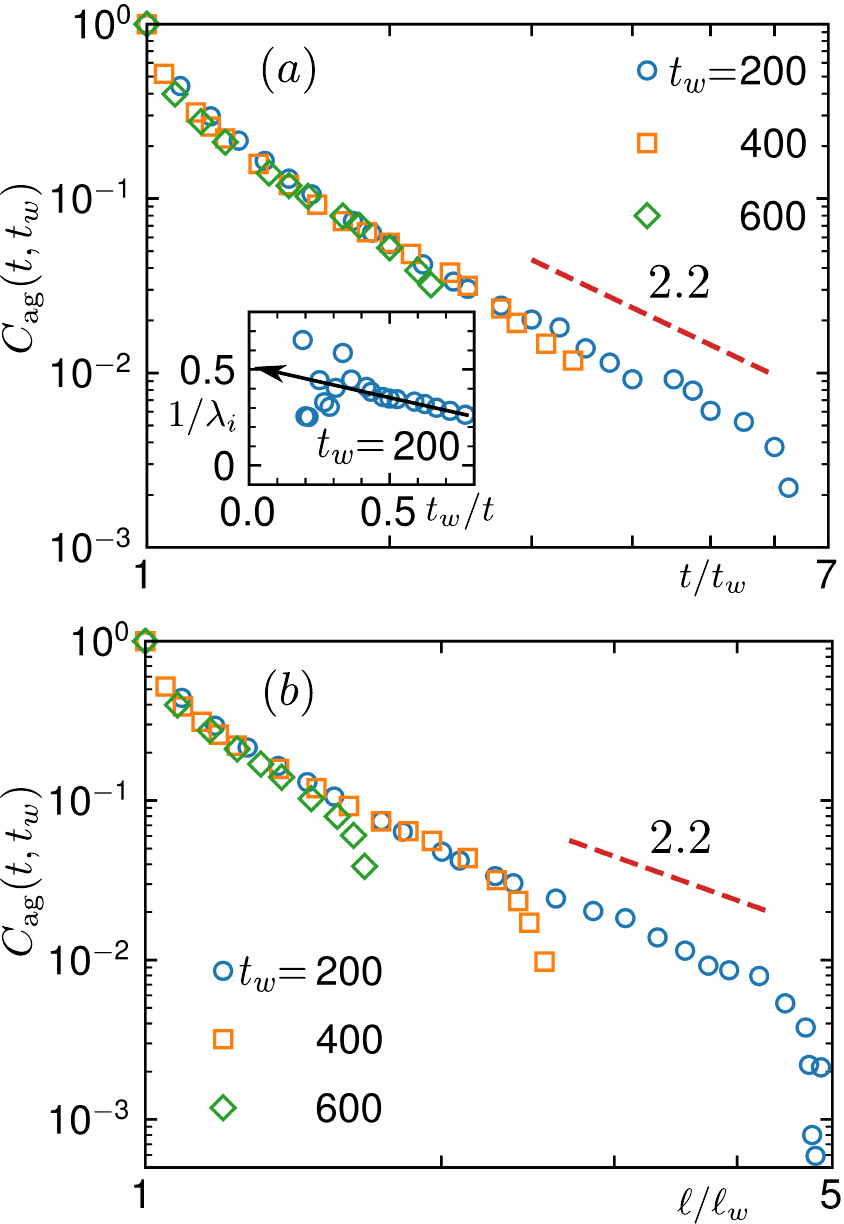}
	\caption{(a)~Plots of $C_{\textrm{ag}}(t,t_w)$ versus $t/t_w$, on a double-log scale, for three values of $t_w$. Inset: Plot of inverse of the (running averaged) instantaneous exponent, $\lambda_i$, versus $t_w/t$, for $t_w=200$. The arrow-headed line here is a guide to the eye.
		(b)~Same sets of data as in the main frame of (a) are plotted versus $\ell/\ell_w$. The dashed lines in (a) and (b) represent power-law with exponent value mentioned in the figure.
	}
	\label{fig:cagscl}
\end{figure}

Motivated by this evidence of aging, we check for the presence of scaling in Fig.~\ref{fig:cagscl}(a), like in the passive systems~\cite{dfisher1988}. There, on a double-log scale, $C_{\textrm{ag}}(t,t_w)$ is plotted versus $t/t_w$. Data for different $t_w$ values nicely collapse on top of each other. This further extends~\cite{das2017} the domain of nonequilibrium scaling phenomena to systems of particles with self-propulsion. Such observation is interesting, considering the fact that in active matter systems the approach of a system is not towards equilibrium, as opposed to the passive case. The scaling function, for large $t/t_w$, seems to have a \mydoubleq{pure} power-law behavior. The dashed line in the figure has a power-law exponent $2.2$, with which the simulation data appear reasonably consistent.
\begin{figure}[!]
	\centering
	\includegraphics*{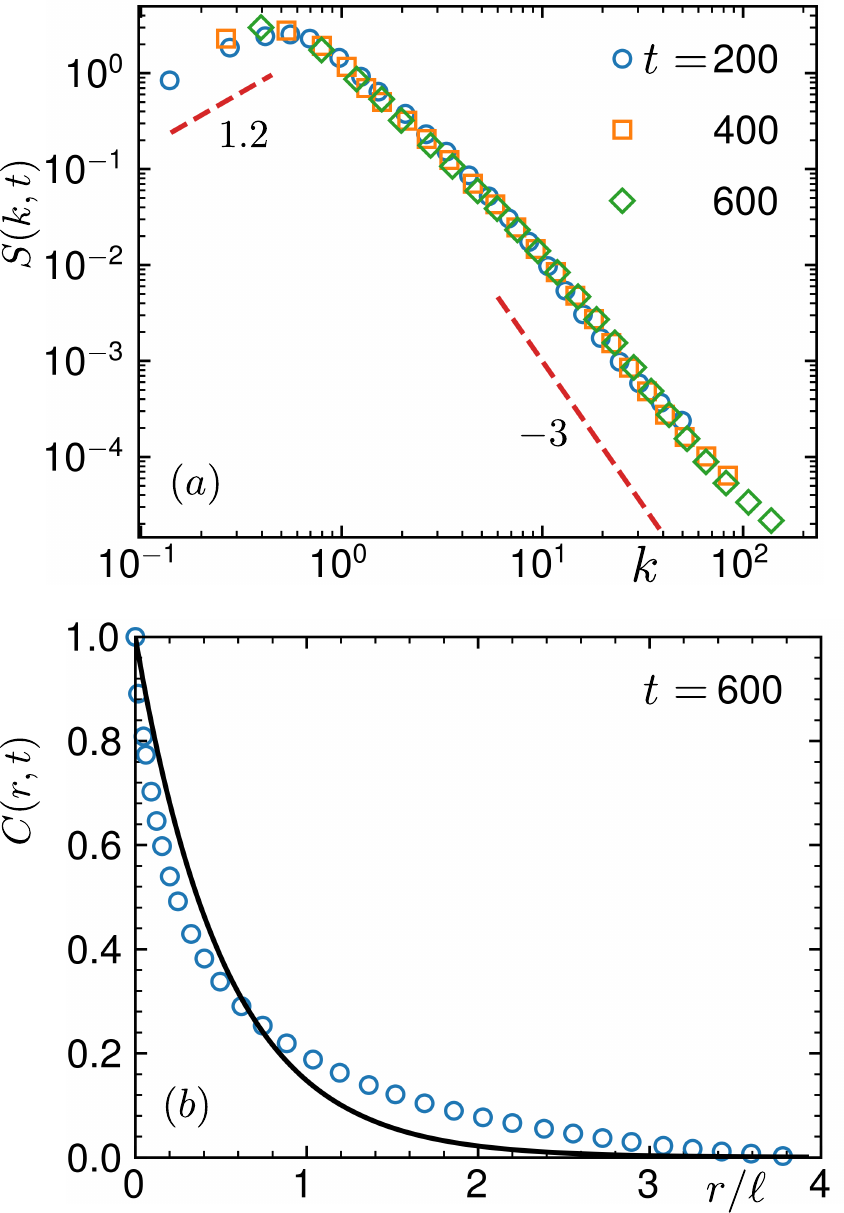}
	\caption{(a)~Scaling plots of structure factor: $\ell^{-d}S(k,t)$ is plotted versus $k\ell$, on a log-log scale. Data from three different times, as mentioned, are presented. Dashed lines indicate power-laws with the exponents noted beside them.
		(b)~Plot of $C(r,t)$, versus $r/\ell$, for $t=600$. The continuous line is a fit to an exponential form [see text for details].
	}
	\label{fig:sclsf}
\end{figure}

For more accurate estimation of the exponent, given that a (upward) bending is present here as well, in the inset of this figure we show a plot of the (inverse of the) instantaneous exponent~\cite{midya2015},
\begin{equation}
\lambda_i=-\frac{d \ln C_{\textrm{ag}}(t,t_w)}{d \ln (t/t_w)},
\end{equation}
versus $t_w/t$. Once again, it is extremely difficult to acquire high quality data for $\lambda_i$, like $\alpha_i$. Nevertheless, reasonably good statistics helped arrive at a meaningful conclusion. The data exhibit convergence towards $\lambda=2.0$. Given that $\alpha \simeq 1$, we have $\lambda \simeq 2.0$ [cf. Eqs.~\eqref{eq:growth} and \eqref{eq:lambda}]. For the sake of completeness, in Fig~\ref{fig:cagscl}(b), we, nevertheless, present the same data sets as a function of $\ell/\ell_w$. Nice collapse of data can again be appreciated. This value of $\lambda$ is much smaller than the corresponding number for diffusive domain growth in passive solid binary mixtures~\cite{midya2015}. Partial reason can be attributed to the morphological differences that we have discussed above. More quantitative picture on this will be provided below. The other important point we like to mention here is the following. The results are consistent with our general observation -- if the dimension is same, value of $\lambda$ is smaller for faster growth~\cite{corberi2019,roy2019}.
 
Next we check whether the obtained value of $\lambda$ satisfies the bound of Eq.~\eqref{eq:yrd}. This analysis will be useful in meaningful comparison of the obtained numbers with those for the same model in $d=3$. Note that the values of $\alpha$ and $\lambda$ in $d=2$ and $3$ are very close~\cite{das2017}.

In Fig.~\ref{fig:sclsf}(a), on a double-log scale, we present scaling plots for $S(k,t)$, using data from three different times. In this case, the expected scaling form~\cite{bray1995} is 
\begin{equation}
S(k,t) \equiv \ell^d \tilde{S}(k\ell).
\end{equation}
Nice collapse of all the data sets on a master curve confirms this form, suggesting again that in spite of the directed motion of the active particles the emergent pattern is self-similar. The dashed lines in Fig.~\ref{fig:sclsf}(a) represent power-laws. Evidently, $S(k,t) \sim k^{1.2}$, in the small $k$ regime. This sets the lower bound of $\lambda$ at $1.6$ [cf. Eq.~\eqref{eq:yrd}]. (Note that for passive systems at this density one expects~\cite{yeung1988} $\beta=4$.) Thus, $\lambda=2.0$ satisfies the bound. At large $k$ (or short distance), rapid power-law decay is reasonably consistent with the Porod-tail~\cite{porod1951,bray1991} behavior $k^{-3}$, expected for scalar order parameter in $d=2$. Minor deviation that is seen can be due to the surface fractality~\cite{shrivastav2014}, which we discussed above.  It is expected that for $d_f<d$, the exponent for the large $k$ decay of the structure factor gets modified to a value smaller than the standard expectation for Porod law~\cite{shrivastav2014}. 

Clearly the small $k$ behavior of the structure factor with $\beta=1.2$ is different from that~\cite{das2017} in $d=3$. To clarify the structural difference between $d=2$ and $3$ cases further, in Fig.~\ref{fig:sclsf}(b) we have shown $C(r,t)$ as a function of $r/\ell$, from $t=600$. The continuous line there is a fit to the form 
\begin{equation}
C(r,t)=\exp\bigg(-\frac{r}{\ell}\bigg),
\end{equation}
that describes the correlation in $d=3$ nicely~\cite{das2017}. The mismatch here again suggests morphological disagreement between the two dimensions. The agreement in the values of $\alpha$ and $\lambda$ may thus be accidental. Furthermore, we believe that the high density phase in the $d=3$ case was liquid~\cite{das2017}. From the calculation of the structure factor during the period of evolution, in this case, we did not observe any Bragg peak in $d=3$.
\begin{figure}[!]
	\centering
	\includegraphics*{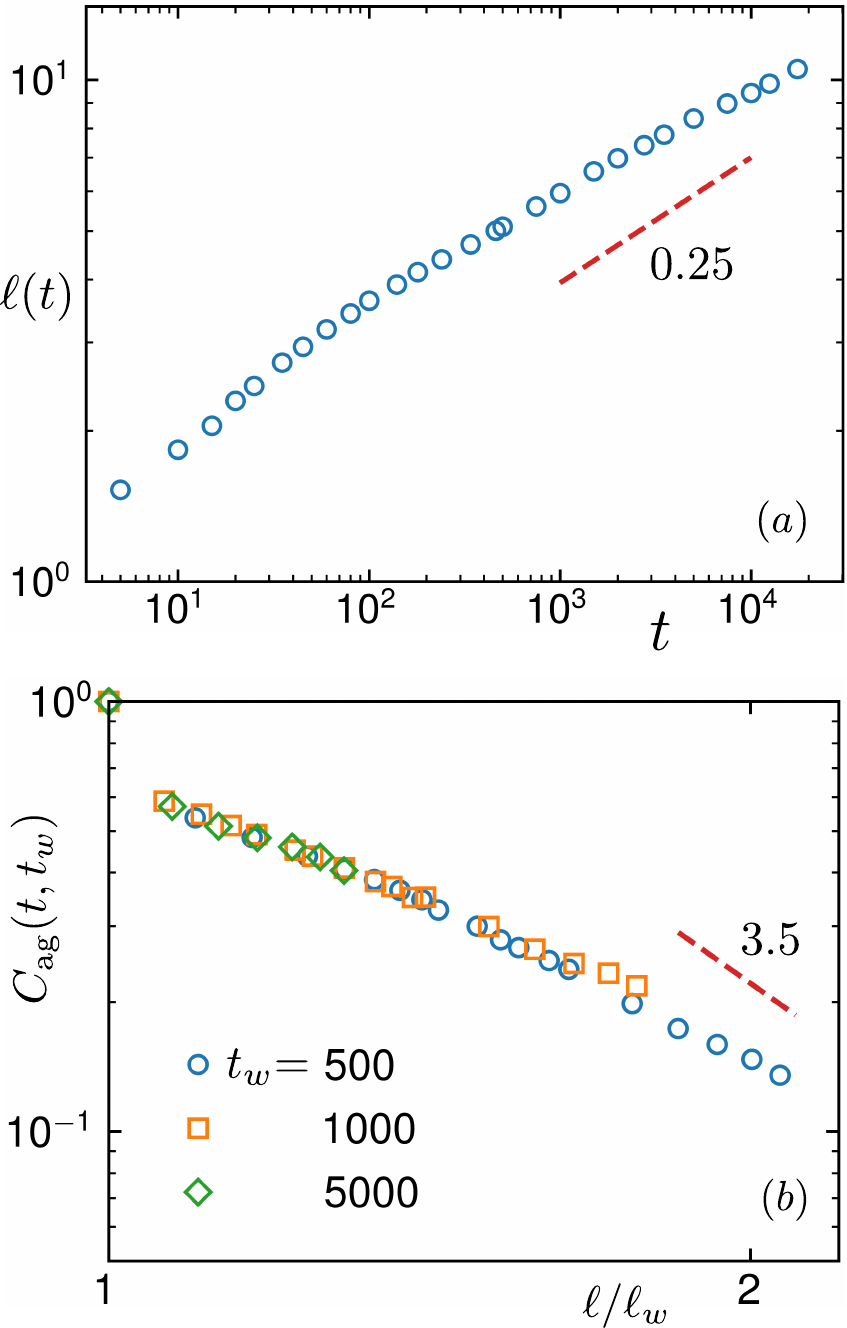}
	\caption{(a)~Plot of $\ell(t)$ versus $t$ for the passive case, i.e., for $f_A=0$. 
	(b)~Scaling plot of $C_{\textrm{ag}}(t,t_w)$ versus $\ell/\ell_w$ for $f_A=0$. Dashed lines represent power-laws with exponents mentioned beside. Here the system size is $L=128$. These results are presented after averaging over ten independent initial configurations.
	}
	\label{fig:pasv}
\end{figure}
\begin{figure}[!]
	\centering
	\includegraphics*{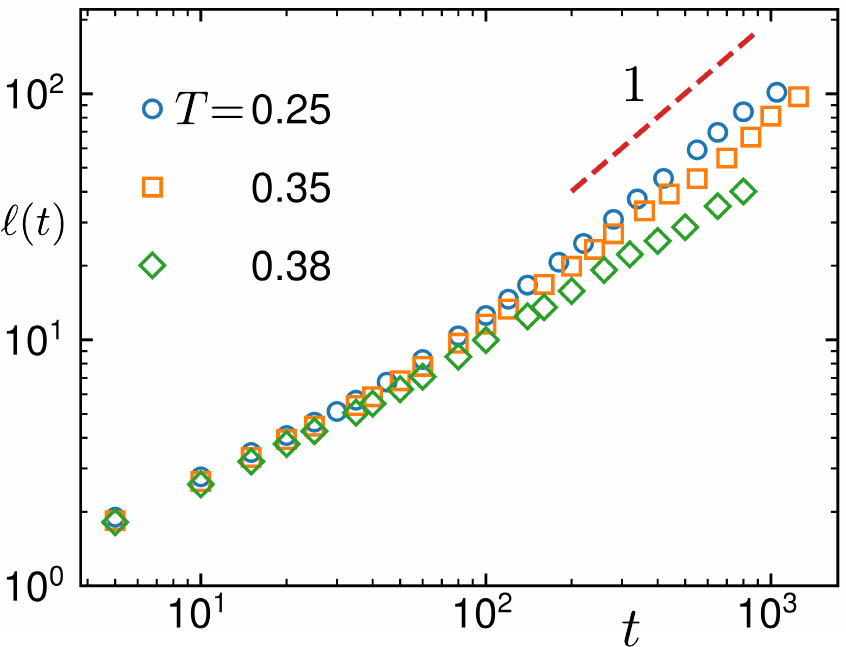}
	\caption{Comparative plots of $\ell$ versus $t$, for the active case, from a few different temperatures. The dashed line represents a power-law with exponent equaling unity. Results for $T$ different from $0.25$ are presented after averaging over ten independent initial configurations. Region(s) with finite-size effects are not shown.
	}
	\label{fig:actv}
\end{figure}

Next we present results for the passive limit of the model. In Fig.~\ref{fig:pasv}(a) we show the plot of $\ell(t)$ versus $t$, for this case, on a log-log scale. Interestingly, the late time growth exponent appears smaller than ${1/3}$, the Lifshitz-Slyozov value~\cite{lifshitz1961}, expected for growth via diffusive transport that is observed in studies of solid mixtures via KIM~\cite{majumder2010}. Our results suggest that the growth problem for the latter should be revisited with more realistic modeling.

In Fig.~\ref{fig:pasv}(b) we show scaling plots of $C_{\textrm{ag}}(t,t_w)$ for the passive case. Good collapse of data is observed for $t_w$ values ranging over a decade. Consideration of a larger range became possible here due to slow growth. Notice, however,  that the range of $\ell_w$ is smaller than that in the active case. The decay of $C_{\textrm{ag}}(t,t_w)$ here is significantly faster than the active case and comparable with that for KIM~\cite{midya2015} in this dimension. This is in line with the observation that $\lambda$ is higher for slower growth~\cite{roy2019}. For a concrete statement, of course, structural quantities must be accurately analyzed. 

Finally, we move to the discussions of temperature dependence of activity mediated phase separation. In Fig.~\ref{fig:actv} we show comparative plots for growth from different temperatures. It appears, with the increase of $T$ the growth becomes slower. But the asymptotic exponent seems to remain temperature independent. Only the convergence to this value gets delayed with the increase of temperature. Quantitative understanding of this issue requires further attention. Corresponding dependence of the aging property is also an important problem.

\section{Conclusion}\label{sec:conclusion}
We have studied dynamics of clustering in a two-dimensional active matter model. The model incorporates Vicsek-like alignment activity~\cite{vicsek1995} into a system of particles interacting via the (passive) Lennard-Jones potential~\cite{allen1987}. The focus was on quantification of the effects of such activity on aging property~\cite{dfisher1988}, along with morphology and growth, when the high density region is in the solid phase. For the study of aging, we have used the two-time order-parameter autocorrelation function as a probe~\cite{dfisher1988}. Comparative discussions of the presented results with the existing and new results for the passive systems and other similar studies of active matter systems are provided. We have also explored the temperature dependence of coarsening in the considered active matter model.

We observe that the autocorrelation function $C_{\textrm{ag}}(t,t_w)$ exhibits scaling with respect to both $t/t_w$ and $\ell/\ell_w$. Power-law scalings in this and in growth~\cite{das2017} extend the domain of non-equilibrium scaling concepts from passive to active systems more firmly. Since active matter models, like the one considered here, are constructed to understand phenomena in systems of living entities, it would be interesting to have more quantitative experimental investigations  of these aspects in pure biological systems~\cite{chen2012,sokolov2012}. 

The exponent $\lambda$ for the decay of the autocorrelation has been estimated to be $2.0 \pm 0.2$. This number obeys a general lower bound~\cite{yeung1996} which we have confirmed via the analysis of pattern. Smaller value of $\lambda$ than that obtained for passive diffusive phase separation, may have its origin in the differences in morphology and growth between the two cases. Note that the value of $\beta$ is different here than the corresponding passive limit~\cite{yeung1988}. Also, the growth happens to be much faster than in systems without activity~\cite{bray1995,majumder2010}. This is consistent with our (recent) previous observation~\cite{roy2019} that faster the growth, smaller the value of $\lambda$. 

The value of $\alpha$ and $\lambda$ appear close to those for the same model~\cite{das2017} in $d=3$. However, the morphological aspects in the two dimensions have differences. Thus, the matching may be accidental.

The model studied here has Vicsek-like~\cite{vicsek1995} self-propulsion. Though simple, this is being helpful in understanding emergent behavior in many biologically motivated systems. The model, however, neglects the effects of hydrodynamics~\cite{hohenberg1977}. Addition of this and other realistic features can make it experimentally more relevant, e.g., in the context of microswimmers in explicit solvent. 

From technical point of view, investigation of hydrodynamic effects in active matter systems is a challenging task, as far as present state-of-the-art is concerned. Interest in the construction of appropriate methodologies for such purpose is gaining momentum only recently~\cite{ali2004,jiang2015}. A way of implementing hydrodynamics could be via the multi-particle collision dynamics~\cite{malevanets2000,gompper2009} method that has found success in the passive matter context. But simulation of large systems with such methods is difficult. This is because of the fact that the number of solvent particles in these simulations is typically an order of magnitude larger than the number of solute particles.

Universality classes concerning scalings related to morphology, growth and aging are reasonably well established in  phase-separating  passive systems~\cite{bray1995}. Such investigations are important for active matter systems as well. Appropriate understanding along that direction, however, requires huge effort with systematic studies.

\section*{acknowledgments}
The authors acknowledge financial support from Department of Biotechnology, India, via Grant No. LSRET-JNC/SKD/$4539$ and from Science and Engineering Board of the Department of Science and Technology, India, via Grant No. MTR/$2019$/$001585$.

The data that support the findings of this study are available from the  authors upon reasonable request.


\end{document}